\documentclass[twocolumn]{aastex631}
\usepackage{booktabs}
\usepackage{makecell}
\usepackage{hyperref}
\usepackage{threeparttable}
\usepackage{bm}

\begin{document}

\title{Edge-on Low-surface-brightness Galaxy Candidates Detected from SDSS Images Using YOLO}


\author{Yongguang Xing}
\affiliation{School of Mechanical, Electrical and Information Engineering, Shandong University, 180 Wenhua Xilu, Weihai, 264209, Shandong, People's Republic of China}

\author{Zhenping Yi}
\altaffiliation{yizhenping@sdu.edu.cn}
\affiliation{School of Mechanical, Electrical and Information Engineering, Shandong University, 180 Wenhua Xilu, Weihai, 264209, Shandong, People's Republic of China}

\author{Zengxu Liang}
\affiliation{School of Mechanical, Electrical and Information Engineering, Shandong University, 180 Wenhua Xilu, Weihai, 264209, Shandong, People's Republic of China}

\author{Hao Su}
\affiliation{School of Mechanical, Electrical and Information Engineering, Shandong University, 180 Wenhua Xilu, Weihai, 264209, Shandong, People's Republic of China}

\author{Wei Du}
\affiliation{Key Laboratory of Optical Astronomy, National Astronomical Observatories, Chinese Academy of Sciences, 20A Datun Road, Chaoyang District, Beijing 100101, People's Republic of China}

\author{Min He}
\affiliation{Key Laboratory of Optical Astronomy, National Astronomical Observatories, Chinese Academy of Sciences, 20A Datun Road, Chaoyang District, Beijing 100101, People's Republic of China}

\author{Meng Liu}
\affiliation{School of Mechanical, Electrical and Information Engineering, Shandong University, 180 Wenhua Xilu, Weihai, 264209, Shandong, People's Republic of China}

\author{Xiaoming Kong}
\affiliation{School of Mechanical, Electrical and Information Engineering, Shandong University, 180 Wenhua Xilu, Weihai, 264209, Shandong, People's Republic of China}

\author{Yude Bu}
\affiliation{School of Mathematics and Statistics, Shandong University, 180 Wenhua Xilu, Weihai, 264209, Shandong, People's Republic of China}

\author{Hong Wu}
\affiliation{Key Laboratory of Optical Astronomy, National Astronomical Observatories, Chinese Academy of Sciences, 20A Datun Road, Chaoyang District, Beijing 100101, People's Republic of China}

\begin{abstract}

Low-surface-brightness galaxies (LSBGs), fainter members of the galaxy population, are thought to be numerous. However, due to their low surface brightness, the search for a wide-area sample of LSBGs is difficult, which in turn limits our ability to fully understand the formation and evolution of galaxies as well as galaxy relationships. Edge-on LSBGs, due to their unique orientation, offer an excellent opportunity to study galaxy structure and galaxy components. In this work, we utilize the You Only Look Once object detection algorithm to construct an edge-on LSBG detection model by training on 281 edge-on LSBGs in Sloan Digital Sky Survey (SDSS) $gri$-band composite images. This model achieved a recall of 94.64\% and a purity of 95.38\% on the test set. We searched across 938,046 $gri$-band images from SDSS Data Release and found 52,293 candidate LSBGs. To enhance the purity of the candidate LSBGs and reduce contamination, we employed the Deep Support Vector Data Description algorithm to identify anomalies within the candidate samples. Ultimately, we compiled a catalog containing 40,759 edge-on LSBG candidates. This sample has similar characteristics to the training data set, mainly composed of blue edge-on LSBG candidates. The catalog is available online at \url{https://github.com/worldoutside/Edge-on_LSBG}.

\end{abstract}


\section{Introduction} \label{sec:intro}

Low-surface-brightness galaxies (LSBGs) are galaxies whose surface brightness is generally fainter than that of the night sky \citep{mcgaugh1995morphology}. Conventionally, LSBGs are face-on galaxies with  central $B$-band surface brightness between 22 $mag \ arcsec^{-2}$ and 23 $mag \ arcsec^{-2}$ \citep{impey2001high, ceccarelli2012low}, or, alternatively, with central $R$-band surface brightness $\mu_{0,R}$ $>$ 24 $mag \ arcsec^{-2}$ \citep{adami2006deep}. The LSBGs make up a large proportion of the luminosity density in the local Universe \citep{impey1997low}, making their contribution to the Universe indispensable.

Due to the edge-on orientation, edge-on LSBGs provide an excellent opportunity to study the vertical structure of galaxies and galactic components \citep{de1998global, bizyaev2017very}. Studies on edge-on LSBGs help to understand galaxy structures such as the interstellar medium, the stellar bar, the galactic wind, and the HI component \citep[e.g.,][]{gerritsen1999star, pohlen2003evidence, matthews2004optical, matthews2008hi}, and may help shed light on star formation and galaxy evolution \citep[e.g.,][]{adams2011new, diaz2022linking, narayanan2022superthin}. In addition, searching for more edge-on LSBGs can increase the number and diversity of edge-on LSBGs to study unusual properties for special edge-on LSBGs \citep{he2019edge}.

A large number of studies have been conducted in selecting and analyzing edge-on LSBGs. For example, some specific edge-on LSBGs have been analyzed in detail, such as UGC 7321, IC 5249, ESO 342-G017, SDSS J121811.0+465501.2, SDSS J092609.45+334304.1, and NGC 4244 \citep[e.g.][]{matthews1999extraordinary, van2001kinematics, neeser2002detection, matthews2003high, liang2007sdss, pustilnik2010sdss, maclachlan2011stability, sarkar2019flaring}. Moreover, thousands of edge-on LSBG samples have been identified and investigated. \citet{matthews2001co} selected eight nearby, edge-on LSBGs using the National Radio Astronomy Observatory 12 m telescope and made CO observations of them. \citet{bizyaev2004stellar} selected a sample of 11 edge-on galaxies from the faintest surface-brightness galaxies in the Revised Catalog of Flat Galaxies \citep{karachentsev1999revised}. \citet{matthews2005detections} selected 15 late-type edge-on LSBGs with the IRAM 30 m telescope and made CO observations. \citet{2006Edge} selected a sample of 970 edge-on LSBGs from the SDSS DR4 data to study the galactic halo emission. \citet{bergvall2010red} selected a sample of 1510 edge-on LSBGs in the SDSS DR5 database to explain the “red halo” phenomenon. \citet{bizyaev2014catalog} obtained 5747 edge-on galaxies by parameter cutting and visual inspection of SDSS DR7 \citep{abazajian2009seventh}. \citet{du2017long} selected a sample of 12 edge-on LSBGs from the catalog of \citet{bizyaev2014catalog} and made spectral observations of them. \citet{he2020sample} selected a sample of 281 edge-on LSBG candidates from the catalog obtained by crossmatching SDSS DR7 with 40\% of the Arecibo Legacy Fast ALFA survey (ALFALFA; \citealp{giovanelli2007alfalfa}) and analyzed the optical and HI properties of the sample.

In previous studies, edge-on LSBGs have primarily been obtained by selecting a central surface brightness and axis ratio. The process of selecting edge-on LSBG samples often involved complicated steps. In addition, the inclusion of skylight contamination in faint galaxies \citep{du2015low} and the sensitivity of edge-on galaxies to initial parameters of the profile fitting \citep{he2020sample} have further complicated the selection process, requiring more manual intervention. Consequently, it becomes challenging to conduct a large-scale, automated search for edge-on LSBGs across sky surveys.

Some machine learning methods have been applied to identify and classify LSBGs to improve efficiency. Traditional machine-learning methods such as support vector machines (SVMs; \citealp{platt1998sequential}) and random forest \citep{breiman2001random} have been employed for LSBG selection. However, their accuracy in identifying LSBGs is only around 50\%, often necessitating a combination with manual inspection \citep{Greco2018,tanoglidis2021shadows}. The low efficiency of these methods in identifying LSBGs is primarily due to using galaxy parameters as training data. The faint surface brightness and complex morphologies of LSBGs make their features difficult to be accurately extracted, leading to significant recognition errors in the identification process. Fortunately, in recent years, deep learning has made significant advancements, greatly enhancing the image analysis capabilities of machine-learning models. Neural networks, with their powerful feature extraction capabilities, can learn galaxy features directly from images, thereby enhancing the ability to identify LSBGs. For example, by using convolutional neural networks (CNNs; \citealp{lecun1998gradient}) to differentiate between LSBG images and artifact images, the model named DeepShadows achieved an accuracy of 92\% \citep{tanoglidis2021deepshadows}. The premise of using cutout galaxy images to identify LSBGs is that a galaxy list has already been successfully obtained. However, some faint galaxies are challenging to be accurately recognized, especially irregular or peculiar galaxies. Additionally, there may be instances where the components of a large galaxy are mistakenly identified as multiple small galaxies, leading to errors in the subsequent identification of LSBGs. Deep-learning-based object detection provides a potential approach for automatically identifying LSBGs. This technique enables the direct recognition and localization of multiple objects from a large image. For example, \citet{yi2022automatic} developed an automated detection model to mainly identify face-on LSBGs in SDSS images and achieved a detection accuracy of 92\%.

In this study, we aim to detect wide-area edge-on LSBGs from SDSS DR16 \citep{ahumada202016th}. Initially, we selected edge-on LSBG candidates using photometric parameters ($expAB\_g$ $\leq$ 0.3 or $expAB\_r$ $\leq$ 0.3, $\mu_{0,B}$ $\geq$ 22 $mag \ arcsec^{-2}$) and obtained 875,993 candidates. However, by examining a subsample of 500 sources, we found that more than half of the samples do not exhibit the morphology of edge-on LSBGs, but rather dense stellar streams, star wings of bright stars, galaxies of nonelongated shape, or irregular morphology. Obtaining true LSBGs requires time-consuming manual inspection of candidates. In this study, we automate the detection of edge-on LSBGs using both object detection and anomaly detection techniques. The constructed object detection model was utilized to automatically identify edge-on LSBGs from SDSS field images, providing both the classification and location of these galaxies.

The layout of this paper is as follows. Section \ref{sec:data} introduces the training and test samples used for building our object detection model. The development of the detection model is described in Section \ref{sec:building}. In Section sec:searching, we present detection results in SDSS DR16 and the process of purifying the candidates. In Section \ref{sec:discussion}, we introduce the properties of the candidate LSBGs. We summarize and conclude our study in Section \ref{sec:conclusion}.

\section{Data Preparation} \label{sec:data}

\begin{figure}
	\centering
		\includegraphics[scale=0.285]{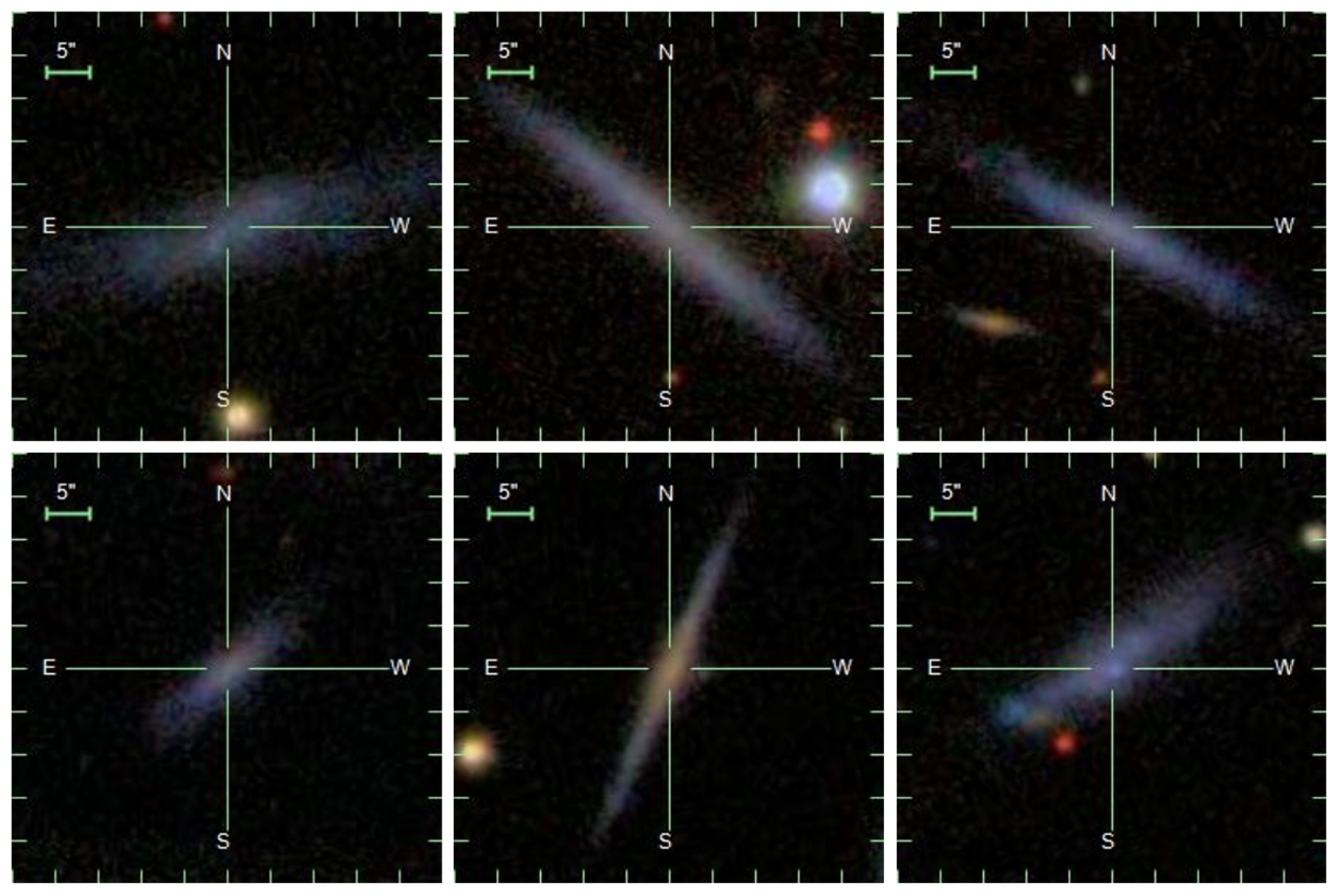}
	\caption{\centering Images of six edge-on LSBG samples in SDSS DR7.}
	\label{FIG:1}
\end{figure}

In this study, we used an edge-on LSBG sample set from \citet{he2020sample} to build an object detection model. This sample set contains 281 edge-on LSBGs that were selected from the crossmatching sample of the 40\% ALFALFA \citep{giovanelli2007alfalfa} catalog and SDSS DR7 \citep{abazajian2009seventh}, with axis ratio $b/a$ $\leq$ $0.3$ in $g$-band or $r$-band and the corrected $B$-band central surface brightness $\mu_{0,B}$ $\geq$ 22.5 $mag \ arcsec^{-2}$. 
Most of the samples are ``blue" galaxies, while a few are ``red" galaxies according to the color ($g-r$; \citealp{bernardi2010galaxy}). In the selection of these edge-on LSBGs, the disparity in surface brightness resulting from the different inclinations of face-on and edge-on galaxies has been corrected, according to \citet{he2020sample}. Figure \ref{FIG:1} shows SDSS images of six edge-on LSBGs in this sample set.

\begin{figure}
	\centering
		\includegraphics[scale=0.285]{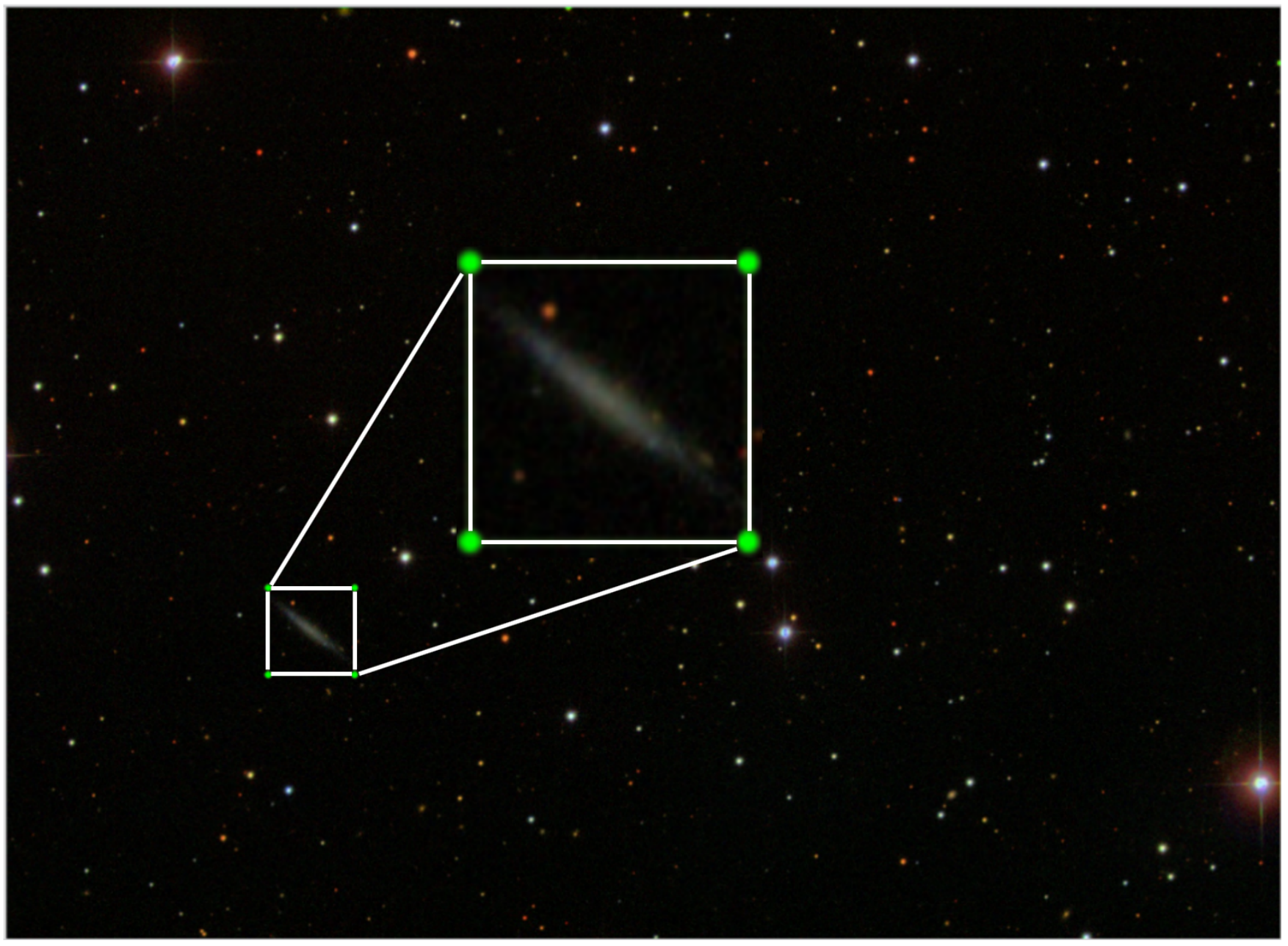}
	\caption{An edge-on LSBG in a labeled bounding box. The image is one of the training samples, which is a $gri$-band composite image of SDSS DR16 obtained from the SAS.}
	\label{FIG:2}
\end{figure}

We obtained 281 $gri$-band composite images of $2048\times1489$ pixels from SDSS DR16, each image containing one edge-on LSBG sample. We labeled these samples using Labelimg to obtain the appropriate size of the bounding box, which is the rectangular box that contains an object. The label parameters include ($x$, $y$, $w$, and $h$), where ($x$, $y$) represents the central coordinate of the target and $w$ and $h$ are the width and height of the bounding box, respectively. Furthermore, we divided the 281 photometric images with a ratio of 8:2, using 225 images as the training set and 56 images as the test set. Figure \ref{FIG:2} shows one edge-on LSBG in an SDSS field image.

\section{Building the Detection Model} \label{sec:building}

\subsection{YOLO}

In this work, we built a model to detect edge-on LSBGs from SDSS field images utilizing the YOLOv5 algorithm, which is one of the You Only Look Once (YOLO) family (\citealp{redmon2016you}; \citet{redmon2017yolo9000, redmon2018yolov3}; \citealp{bochkovskiy2020yolov4}). As a typical representative of one-stage algorithms, the YOLO algorithm directly extracts feature values and has the advantage of detection speed, accuracy, and learning capabilities. The YOLOv5 algorithm combines the ideas of the previous YOLO algorithms and has been innovated in terms of data augmentation, effectively solving the problem of not detecting small targets. In addition, YOLOv5 has a highly desirable speed of training and detection. In this work, we chose the medium-sized network architecture YOLOv5m provided by the Ultralytics YOLOv5 library to build an object detection model for detecting edge-on LSBGs.

The base model we used is the “Medium” variant of the YOLOv5 models, named YOLOv5m. The network structure of YOLOv5m consists of four parts: the input module, which is used to acquire input images and perform data augmentation; the backbone module, which extracts high-, medium-, and low-level features from images; the neck module, which fuses feature information from all levels of the image and extracts large, medium, and small feature maps; and the head module, which applies anchor boxes to the generated feature maps for final detection. The loss function consists of the localization loss $L_{loc}$, classification loss $L_{cls}$, and confidence loss $L_{conf}$. We used complete intersection over union (CIoU) as the localization loss and binary cross-entropy loss as the classification loss. The total loss is a linear sum of the three losses, which is shown in the following equation.

\begin{equation}
    Loss = \lambda_{1}L_{loc} + \lambda_{2}L_{cls} + \lambda_{3}L_{conf}
    \label{eq0}
\end{equation}
where $\lambda_{1}$, $\lambda_{2}$, $\lambda_{3}$ are weighting factors assigned to the localization loss, classification loss, and confidence loss. In our experiments, $\lambda_{1}$, $\lambda_{2}$, $\lambda_{3}$ were kept to be default values of 0.05, 0.5, and 1.0, respectively.

\subsection{Model Training}

In the training process of the model, the input image was resized to $640\times640$ pixels. The batch size was set to 12. To optimize the model parameters, the Adam optimizer \citep{kingma2014adam} was adopted. Moreover, the parameters image translation, image scale and image mosaic were set to 0.1, 0.5, and 1.0, respectively. The other hyperparameters, such as the initial learning rate, IoU training threshold, etc., were kept as default values.

After 150 epochs of training, the localization loss and confidence loss of the model achieved a smooth convergence. Finally, we got an object detection model that can be used to search for edge-on LSBGs in SDSS DR16 composite images.

\subsection{Performance Evaluation}

We used recall and purity to evaluate the performance of our detection model on the test data set. The two measures are calculated as follows:

\begin{equation}
    recall=\frac{TP}{N}
    \label{eq1}
\end{equation}

\begin{equation}
    purity=\frac{TP+CP}{TP+FP}
    \label{eq2}
\end{equation}
where TP (true positives) is the number of detected samples among the labeled LSBGs, N is the total number of labeled samples, FP (false positives) is the number of newly detected objects, CP (check positives) is the number of candidate LSBGs confirmed by checking their shape and central surface brightness (SDSS parameters were used to calculate their $B$-band central surface brightness according to \citet{he2020sample}; a galaxy with $B$-band central surface brightness greater than 22 $mag \ arcsec^{-2}$ is determined as a candidate LSBG).

\subsection{Model Testing}

Our model outputs the center position of the detected object, the width and height of the bounding box, and the confidence. The confidence is the probability of an edge-on LSBG in the bounding box. In our experiments, the default confidence threshold of 0.25 was used in model testing; that is, the object with a confidence lower than 0.25 would not be retained.

The built model detected edge-on LSBGs from SDSS images in the test set. From the 56 SDSS field images of the test set, we have detected 90 sources, which include all 56 previously labeled edge-on LSBGs and 34 newly detected sources. We performed a visual inspection of the newly detected sources to see if they were galaxies. Furthermore, their $B$-band central surface brightness is calculated using the SDSS parameters, and the cutting of $\mu_{0,B}$ $\geq$ 22 $mag \ arcsec^{-2}$ is implemented. Finally, 25 of the 34 candidates are considered to be candidate LSBGs.

\begin{figure}
	\centering
		\includegraphics[scale=0.43]{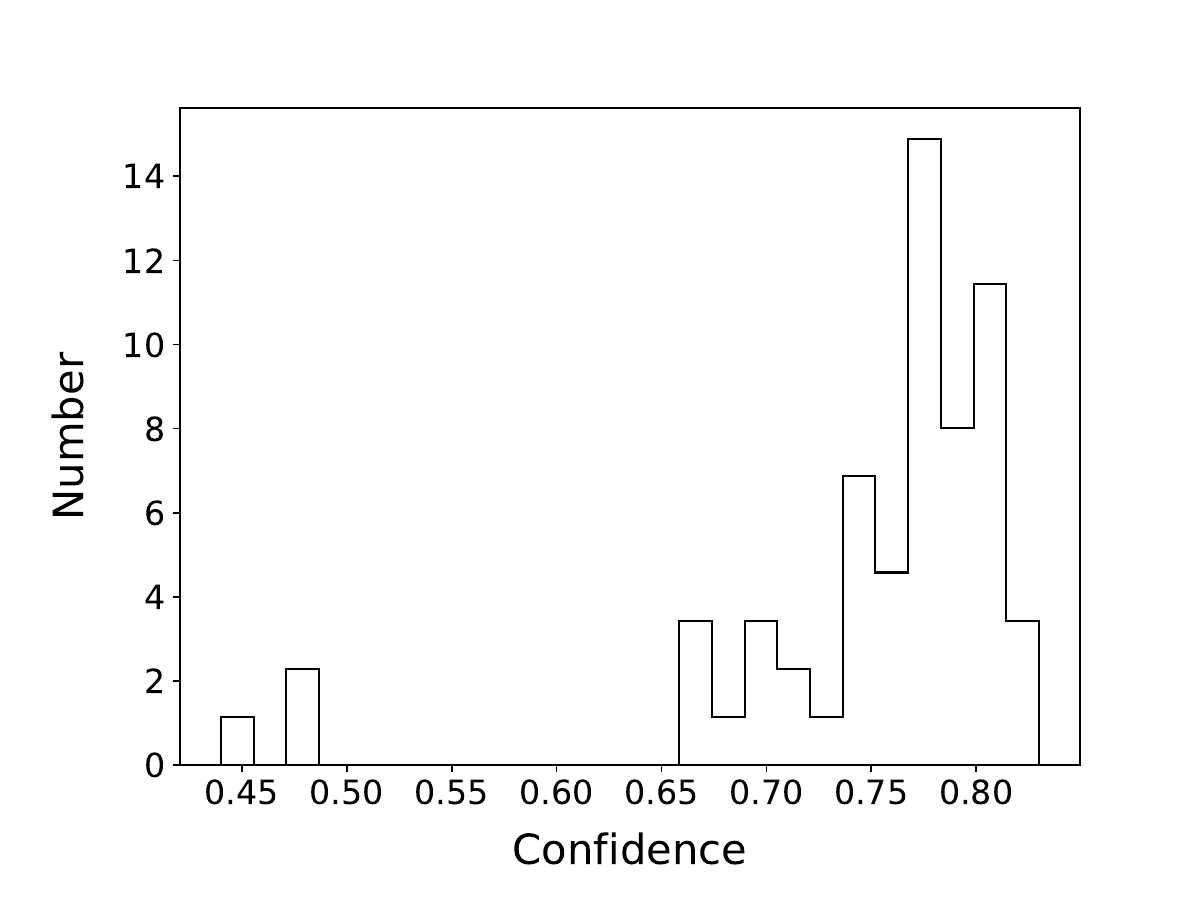}
	\caption{Distribution of confidence for all the labeled edge-on LSBGs in the test set detected by the YOLOv5m model.}
	\label{FIG:3}
\end{figure}

Figure \ref{FIG:3} shows the confidence distribution of the 56 correctly detected samples from the test set. Among them, only three had significantly lower confidence (the values are 0.48, 0.48, and 0.44, respectively) because of their abnormal shapes, while the remaining 53 samples had confidence values greater than 0.65.

We attempted to enhance the purity of the detected samples by increasing the confidence threshold. We set the confidence thresholds to 0.45 and 0.65, resulting in purity of 93.33\% and 95.38\% and recall rates of 98.21\% and 94.64\%. The detailed test results of the model at three different confidence thresholds are shown in Table \ref{tbl1}. The test results suggest that as the the confidence threshold increases, the purity of the detected samples increases, but the recall rate decreases. Trading off recall and purity, we chose a confidence threshold of 0.65 for the subsequent edge-on LSBG detection.

\begin{table}[]
 \centering
    \caption{Test Results of Our Detection Model at Three Confidence Thresholds.}
 \label{tbl1}

\begin{tabular}{cccccc}
\hline
\hline
Confidence & TP$^{a}$    & FP$^{b}$    & CP$^{c}$    & Recall  & Purity      \\ \hline
0.25       & 56          & 34          & 25          & 100\%   & 90\%        \\
0.45       & 55          & 20          & 15          & 98.21\% & 93.33\%     \\
0.65       & 53          & 12          & 9           & 94.64\% & 95.38\%     \\ \hline
\end{tabular}
\begin{threeparttable}
\begin{tablenotes}
    \footnotesize
    \item[a] TP (true positives) is the number of detected samples that are already labeled in the training data set.
    \item[b] FP (false positives) is the number of newly detected candidates.
    \item[c] CP (check positives) is the number of newly detected candidates that are confirmed as candidate LSBGs by checking the shape and central surface brightness in B-band.
  \end{tablenotes}
\end{threeparttable}
\end{table}

\section{Searching for Edge-on LSBGs from SDSS DR16} \label{sec:searching}

Here we describe the pipeline used to search for edge-on LSBGs from SDSS DR16. First, we use the established object detection model to search for edge-on LSBGs across all composite images from SDSS DR16. In addition, we trained an anomaly detection model to help remove contamination among the candidates, thereby improving the automation of the sample purification process.

\subsection{Data Preparation}

We obtained the $gri$-band composite images from the SDSS DR16 Science Archive Server (SAS), a total of 938,046 images, each with a size of $2048\times1489$ pixels. These images have been preprocessed with flat field, bias corrections, bad pixel corrections, and sky subtraction. 

\subsection{Results}

The detection process was executed on a platform equipped with an NVIDIA GTX 1660 GPU. Each image's detection took approximately 0.023 s, resulting in a cumulative processing time of 18.91 hr for all 938,046 images. Within this dataset, our model identified 52,293 candidate LSBGs. The model predicted their central coordinates, bounding box dimensions (width and height), and assigned a confidence level. As an illustration, Figure 4 displays an SDSS image where our model identified two candidate LSBGs. 

\begin{figure*}
	\centering
		\includegraphics[scale=0.5]{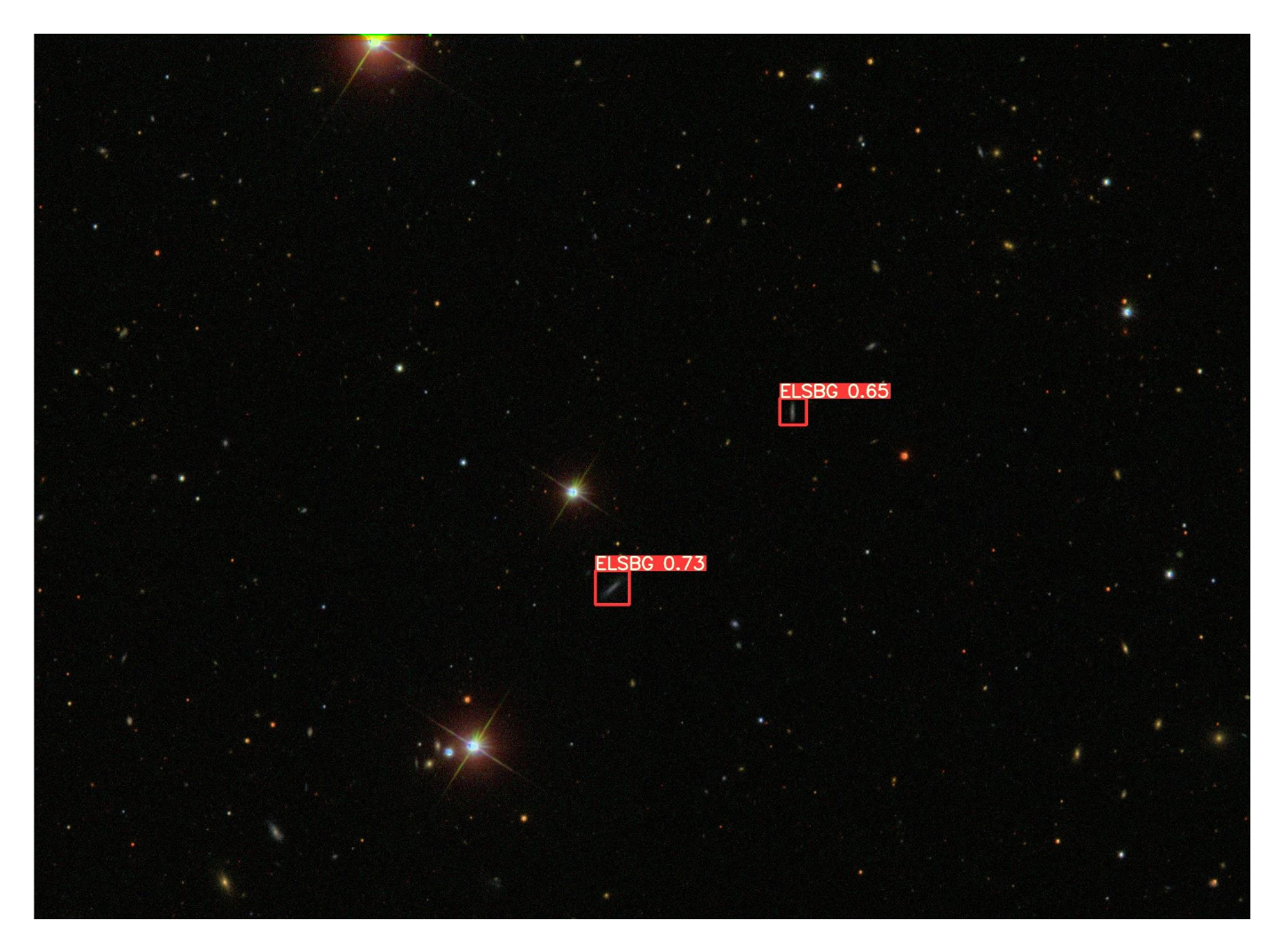}
	\caption{Two candidates detected by the model in an SDSS composite image, with the confidence of the candidates predicted by our model.}
	\label{FIG:4}
\end{figure*}

\subsection{Purify the Samples Using Anomaly Detection}

Through visual inspection of a random subsample of the detected candidate LSBGs, we found that there were a small number of candidates that had similar characteristics to edge-on LSBGs, such as some star wings and thin artifacts. Therefore, to further improve the purity of the sample of candidate LSBGs, we next tried to identify the anomalous candidates.

Anomaly detection \citep{chandola2009anomaly} is a data analysis technology used for identifying and detecting anomalous samples in a data set. As one of the anomaly detection algorithms based on deep learning, Deep Support Vector Data Description (Deep-SVDD; \citealp{ruff2018deep}) can be used to process complex data such as images and directly identify a small percentage of anomalous image samples. Deep-SVDD uses deep neural networks to model complex data distributions and can learn a compact representation of the normal class data, for solving one-class classification problems. One-Class Deep-SVDD finds a hypersphere of minimum volume with center $c$ and contracts the sphere by minimizing the mean distance of all data representations to the center. To achieve this goal, the neural network must extract the common factors of variation. For some input space $X\subseteq R^{d}$ and output space $F\subseteq R^{p}$, $\phi (\cdot;w):$ $X \longrightarrow F$ is a neural network with $L\in N $ hidden layers and set of weights  ${W}=\{W^1,...,W^L\}$, where $W^{l}$ are the weights of layer $l\in \{1,...,L\}$. Given some training data $D_n=\{x_1,...,x_n\}$ on $X$, the loss function of One-Class Deep-SVDD objective is defined as follows:

\begin{equation}
    \label{eq4}
    \min_{w} \frac{1}{n} \sum_{i=1}^{n} \Vert\phi (x_{i};w ) -c\Vert^2 +\frac{\lambda}{2} \sum_{l=1}^{L}\Vert W^{l}\Vert_{F}^{2} 
\end{equation}
The first term of the loss function employs a quadratic loss for penalizing the distance of every network representation $\phi (x_{i};w )$ to center $c\in F$. The second term is a network weight decay regularizer with hyperparameter $\lambda>$ 0, where $ \parallel \cdot \parallel_{F}$ denotes the Frobenius norm. During testing, the input data points are mapped to the latent space using the trained neural network. The distance between each data point and the center of the hypersphere is calculated, known as the anomaly score, and if the anomaly score is greater than a threshold, the data point is considered an anomaly. The equation for calculating the anomaly score is as follows:

\begin{equation}
    s(x) = {\Vert \phi(x; w^*)-c \Vert}^2
    \label{eq5}
\end{equation}
where $x$ represents the test data, $w^*$ represents the network parameters of a trained one-class classification model, $\phi(x; W^*)$ represents the network representations, and $c$ represents the center of the hypersphere.

We used the previously mentioned data set of 281 edge-on LSBGs as training samples to train the anomaly detection model. These samples were obtained in the form of composite images with dimensions of $250\times250$ pixels from the SDSS SkyServer. As part of the preprocessing step, we conducted image scaling and applied random flipping.
For building the anomaly detection model, we first utilized a deep convolutional autoencoder (DCAE; \citealp{masci2011stacked}) to initialize the network weights and obtain the hypersphere center $c$, which is set to the mean of the mapped data after performing pretraining. The DCAE consists of an encoder and a decoder. The encoder compresses the input data to learn informative features, while the decoder decompresses the learned representations to reconstruct the original input. Here, the DCAE encoder has the same architectures as the Deep-SVDD network. A LeNet-type CNN is used in the Deep-SVDD network, where each convolutional module consists of a convolutional layer followed by leaky ReLU activations and $2\times2$ max pooling. Three CNN modules are used, including $32\times(5\times 5\times 3)$ filters, $64\times(5\times 5\times 3)$ filters, and $128\times(5\times 5\times 3)$ filters, followed by a  dense layer of 128 units (see \citealp{ruff2018deep} for details \footnote{Deep-SVDD code: \url{https://github.com/lukasruff/Deep-SVDD}}).
The pretraining used the normal samples of edge-on LSBGs. After pretraining the DCAE, we obtained the hypersphere center $c$. Then we trained the LeNet-type CNN using one-class classification loss, with initial weights from the trained DCAE to obtain the final model. We trained 500 epochs for both the DCAE and LeNet-type CNN. After completing the training process, we obtained a model capable of assigning anomaly scores to candidate images. Eventually, the detection model was applied to 52,293 candidate LSBGs and their anomaly scores were obtained. As a result, their scores were distributed in the range of 0.01-19.21. Most of the candidates had small anomaly scores close to 0.01, while a few had relatively significant scores.

\begin{figure}
	\centering
		\includegraphics[scale=0.5]{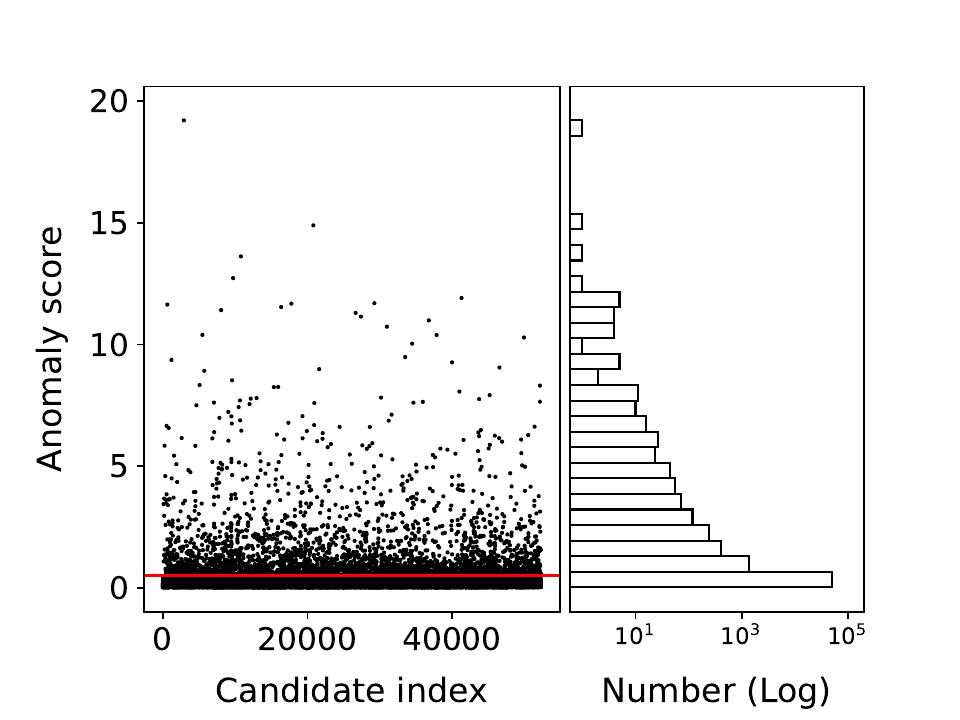}
	\caption{The distribution of the anomaly scores. The left panel shows a scatter plot, and the right panel is a histogram plot. The red line in the scatter plot represents the upper limit of 0.49.}
	\label{FIG:5}
\end{figure}

To establish an appropriate threshold for anomaly scores, we employed the box-plot method \citep{tukey1977exploratory}, which visually illustrates the distribution, dispersion, and skewness characteristics of numerical data using quartiles \citep{dutoit2012graphical}. To determine the threshold for the anomaly scores, we set the upper limit in the box plot to Q3+3IQR (inter-quartile range, where IQR=Q3-Q1, and the factor 3 is used to identify extreme outliers), where Q1 is the lower quartile, Q3 is the upper quartile, and for this case they were calculated as 0.042594 and 0.154938, respectively. Consequently, we established an upper-limit value of 0.49. 
Furthermore, we computed the anomaly scores for all 281 training samples, and all of them fell below the upper-limit line at y = 0.49. Figure \ref{FIG:5} displays the distribution of the anomaly scores for the candidate LSBGs.

\begin{figure}
	\centering
		\includegraphics[scale=0.465]{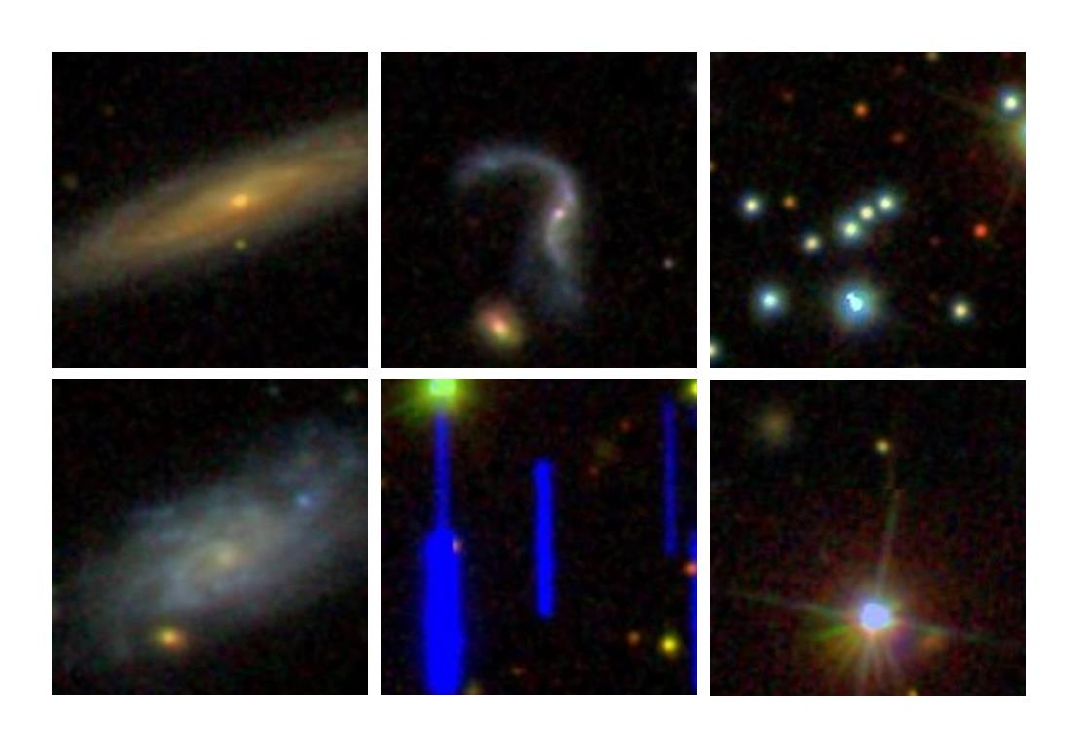}
	\caption{Examples of anomalous sources detected by the model. Their anomaly scores are 0.67, 1.00, 2.12, 2.44, 6.74, and 11.14, respectively.}
	\label{FIG:6}
\end{figure}

There are 3403 candidates located above the upper-limit line (y = 0.49), accounting for about 6.51\% of the total candidate LSBGs. A visual inspection of these candidates revealed that most exhibited  unusual characteristics, including artifacts, galaxies with larger axis ratios, merging galaxies, star wings, and star lines. They were incorrectly identified, either because of their similar shape to edge-on LSBGs or because of the interference from nearby bright stars. Figure \ref{FIG:6} shows six images of these anomalous candidates detected by the anomaly detection model. After a visual inspection, we removed 1981 anomalous sources that did not match the profile of edge-on LSBGs. The remaining 1422 candidates exhibited the characteristic shape of edge-on galaxies, and thus they were retained. Their elevated anomaly scores primarily stemmed from the presence of other celestial objects in proximity, such as bright stars and galaxies. Consequently, 50,312 candidate LSBGs remained.

Following the manual inspection of candidates with outlier scores exceeding 0.49, we conducted random sampling checks on the remaining 48,890 samples with scores below 0.49. These checks were performed through five sets of sampling, each comprising 200 samples, and were designed to detect potential contaminants (our consideration revolves around the straightforward identification of contamination resulting from artifacts, star wings, and similar factors, without taking into account more intricate parameters like surface brightness or axis ratio, which can be challenging for the human eye to discern). Across the five sets, we observed contamination rates ranging from 2.5\% to 0.5\%, which decreased as the outlier scores decreased. The average contamination rate across the five sets stood at approximately 1.6\%.
While our automated process did not entirely eliminate erroneous sources, this 1.6\% contamination rate signifies a significant improvement in purity compared to the results achieved by previous machine-learning methods. For reference, LSBG candidates identified using an SVM classifier exhibited a contamination rate of approximately 47\% \citep{tanoglidis2021deepshadows}.

\section{Discussion} \label{sec:discussion}

\begin{figure}
	\centering
		\includegraphics[scale=0.285]{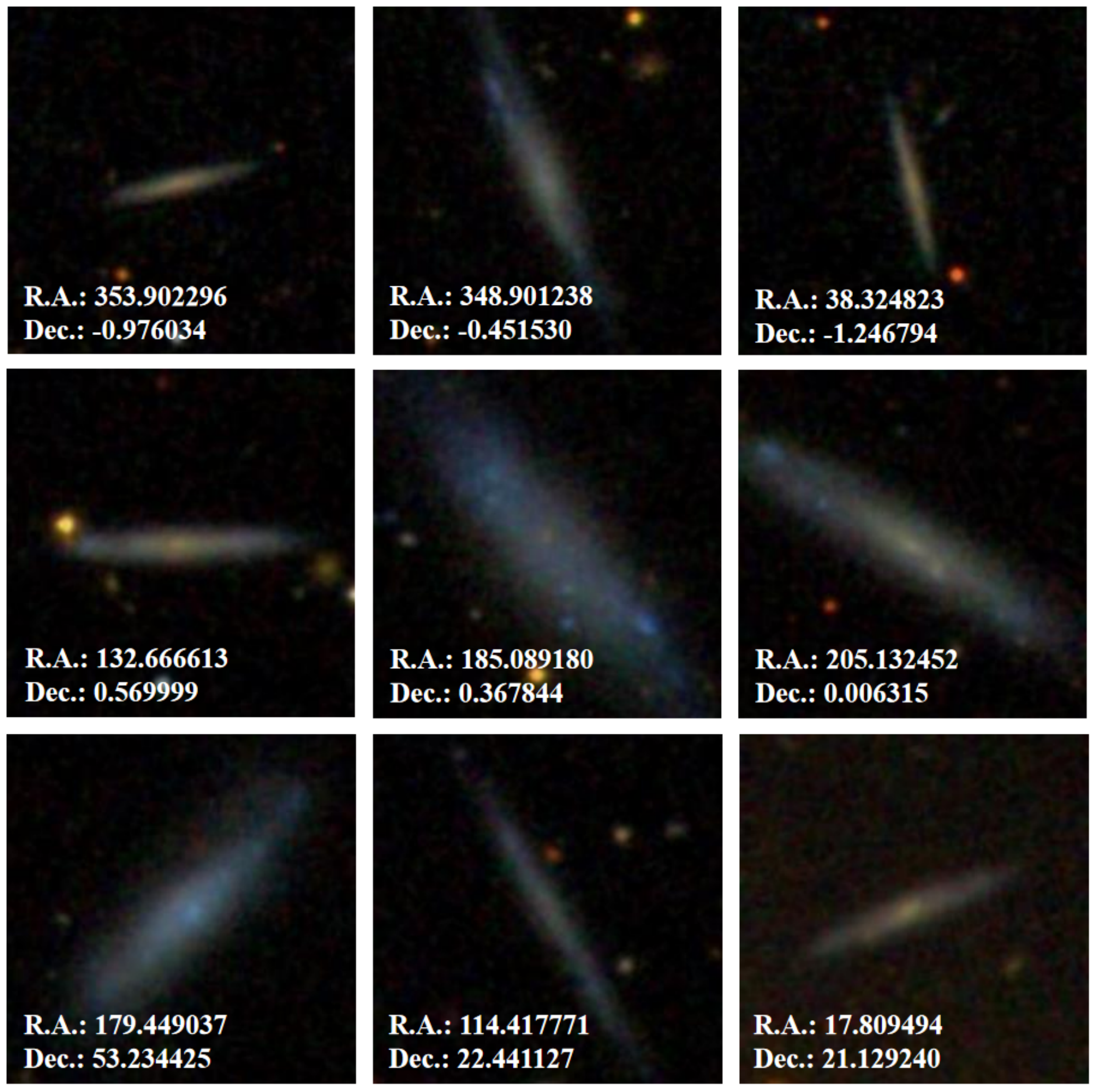}
	\caption{\centering  Six candidate LSBGs identified by our object detection model, which are also included in the galaxy view of SDSS DR16 (the first two rows). And three candidate LSBGs identified by our model, which are not included in the galaxy view of SDSS DR16 (the last row).}
	\label{FIG:7}
\end{figure}

To show the properties of the detected candidate LSBGs, we crossmatched the detected samples with the galaxy catalog of SDSS DR16 to obtain the parameters provided by SDSS. With a search radius of $12^{\prime\prime}$, we got a total of 49,972 matched galaxies. Regarding the remaining 340 unmatched candidate samples, we cannot find the corresponding source in the SDSS DR16 galaxy catalog. Upon inspecting their images, we determined that 201 of them were edge-on galaxies, while the remaining 139 were artifacts and star lines. After excluding these 139 samples, our data set comprised 50,173 edge-on LSBG candidates. Figure \ref{FIG:7} shows the nine edge-on LSBG candidates that we have identified. Among these candidates, the top two rows include six that are listed in the SDSS catalog, whereas the three candidates in the last row were newly identified by our model.

\subsection{Properties of candidate LSBGs}
\begin{figure}
	\centering
		\includegraphics[scale=0.28]{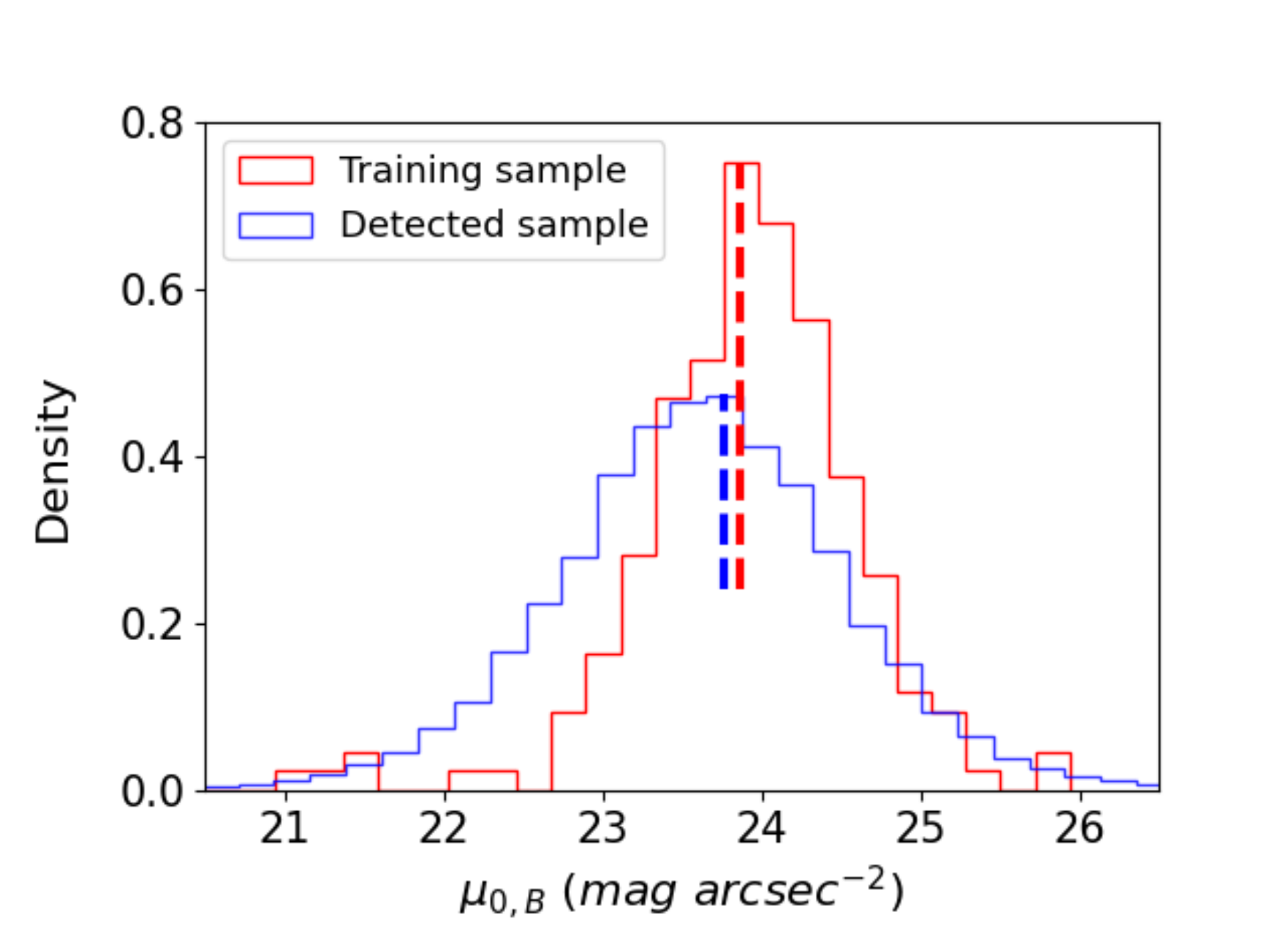}
	\caption{The normalized distribution of $\mu_{0,B}$ of the training edge-on LSBGs and the detected candidate LSBGs. The red and blue vertical lines are marked at the peaks of the training samples and detected samples, respectively.}
	\label{FIG:8}
\end{figure}

For the 49,972 matched edge-on LSBG candidates, we obtained their photometric parameters from SDSS SkyServer. To get the B-band central surface brightness $\mu_{0,B}$, we removed 3136 candidates with invalid parameters ($expRad\_g$ or $expRad\_r$ less than $1^{\prime\prime}$, redshift $z$=-9999), leaving 46,836 candidate LSBGs. Then we obtained their $\mu_{0,B}$ following the calculation method of \citet{he2020sample}, in which $\mu_{0,B}$ is corrected for the inclination effect of edge-on LSBGs. Figure \ref{FIG:8} shows the distribution of number density of galaxies against the $B$-band central surface brightness $\mu_{0,B}$ for the detected sample (in blue). For comparison, the density distribution of $\mu_{0,B}$ of the training samples (in red) is also presented in Figure \ref{FIG:8}.

As can be seen from Figure \ref{FIG:8}, the $\mu_{0,B}$ of the detected candidate LSBGs is mainly distributed in the range of 21-26 $mag \ arcsec^{-2}$, in agreement with that of the training samples. The peak value of the $\mu_{0,B}$ of the detected samples is 23.76 $mag \ arcsec^{-2}$, about 0.11 $mag \ arcsec^{-2}$ lower than the peak value of the central surface-brightness distribution of the training samples. One possible reason for the slight deviation in surface brightness is that surface brightness is essentially a challenging learning feature, as it involves many related factors. Another influencing factor is our limited number of training samples.

\begin{figure}
	\centering
		\includegraphics[scale=0.5]{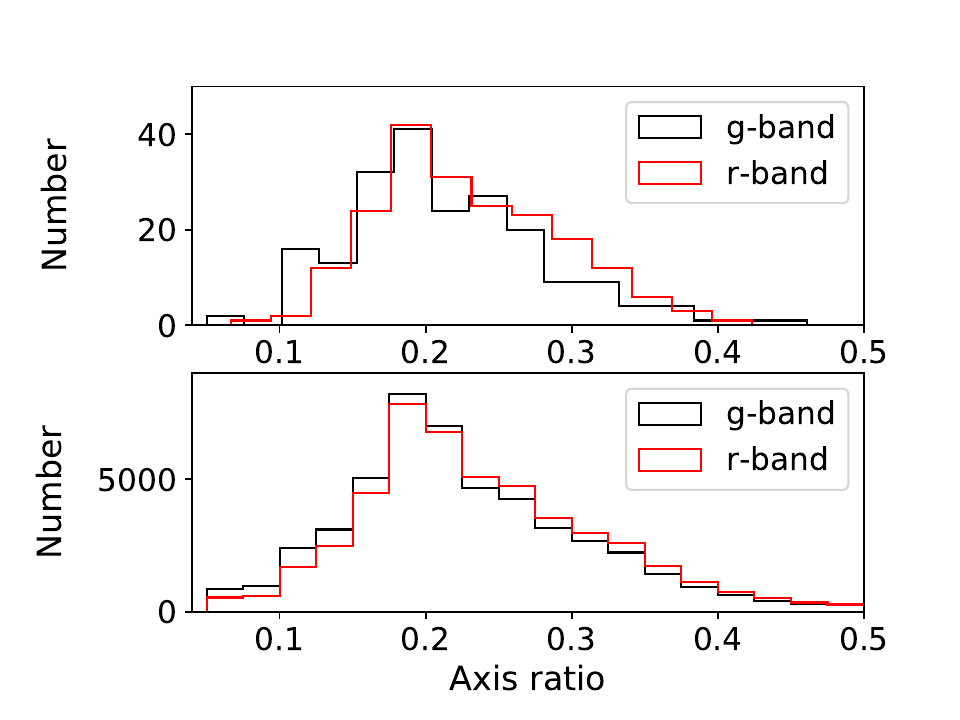}
	\caption{Distribution of axis ratio in $g$ band and $r$ band of training edge-on LSBGs (top panel) and identified candidate LSBGs (bottom panel).}
	\label{FIG:9}
\end{figure}

Additionally, we present the axis ratio distributions of the training samples and detected samples in the $g$ band and $r$ band, determined based on the SDSS photometric parameters $expAB\_g$ and $expAB\_r$, as depicted in Figure \ref{FIG:9}. It is evident that the axis ratio distributions of the detected samples in the $g$ band and $r$ band have good consistency with those of the training samples, which indicates that our detection model has extracted the slender shape features of edge-on galaxies well, thereby enabling the effective detection of candidate LSBGs.

\begin{figure*}
\hspace{-0.5cm}
	\centering
		\includegraphics[scale=0.55]{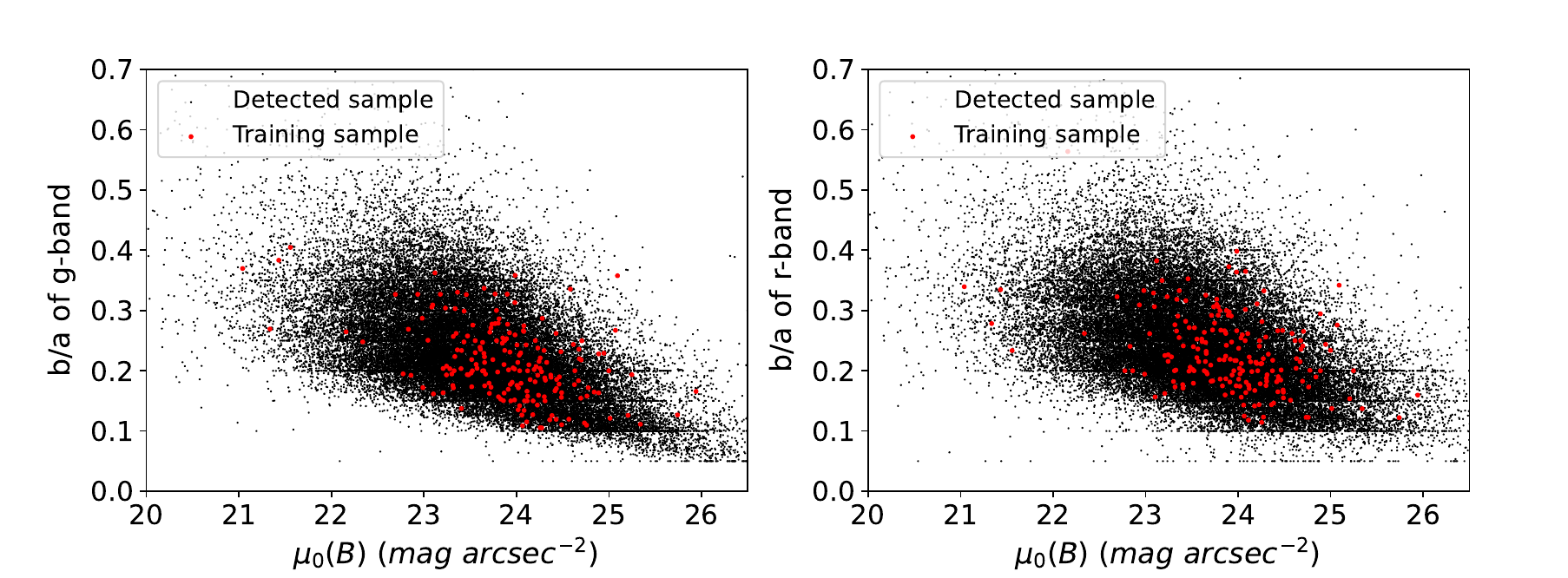}
	\caption{Distribution of axis ratio vs. $B$-band central surface brightness for training and detected samples in $g$ band (left panel) and $r$ band (right panel).}
	\label{FIG:10}
\end{figure*}

The relation between the axis ratio and the central surface brightness is shown in Figure \ref{FIG:10}. Both the identified candidates and the training samples show the same distribution, where as the axis ratio decreases, galaxies with lower surface brightness become detectable, indicating that galaxies with lower axis ratios are easier to detect due to the accumulation of luminosity. Additionally, the detected galaxies exhibit a broader distribution of axis ratios and surface brightness than that of the training sample. Some detected sources fall beyond the range of axis ratios and surface brightness for edge-on LSBGs. This indicates that the model is capable of detecting relatively faint galaxies with slender shapes, but lacks the ability to accurately distinguish sources with axis ratio and surface brightness near the threshold boundaries. Fewer negative samples of axis ratio and surface brightness near the cutting thresholds may have resulted in insufficient learning of the identification boundaries.

\begin{figure}
	\centering
		\includegraphics[scale=0.5]{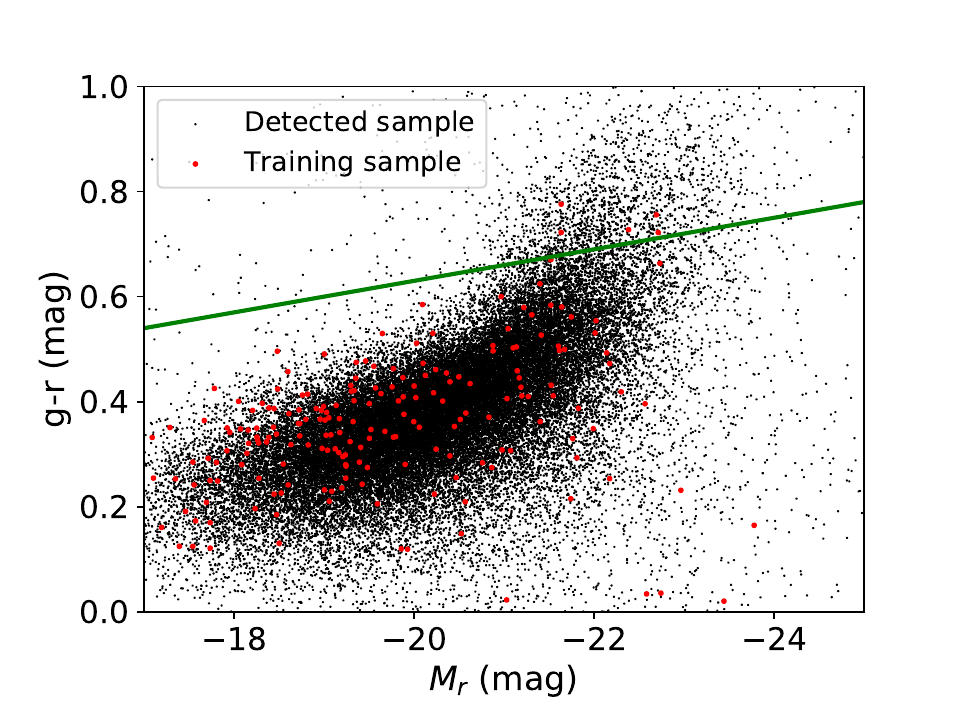}
	\caption{Color–magnitude relation for the detected candidate LSBGs (black) and training edge-on LSBGs (red). The green line is the dividing line between “red”-sequence galaxies and “blue” cloud galaxies \citep{bernardi2010galaxy}.}
	\label{FIG:11}
\end{figure}

The colour–magnitude relation for training samples and detected samples is shown in Figure \ref{FIG:11}. The absolute magnitude $M_r$ and color $(g-r)$ have been corrected according to \citet{he2020sample}, considering the differences in internal extinction and color changes between face-on and edge-on galaxies. The green line is the dividing line between ``red”-sequence galaxies and ``blue” cloud galaxies \citep{bernardi2010galaxy}. ``Red" galaxies lie above this line, while ``blue" galaxies lie below it. Figure \ref{FIG:11} shows that most of our edge-on LSBG candidates are located in the ``blue” region, exhibiting excellent consistency with the training data.

In summary, the properties of the detected candidates are generally consistent with those of the training samples, affirming the effectiveness of our detection model. The identified galaxies exhibit slender shapes and relatively low surface-brightness features, indicating that our detection model has effectively captured the key characteristics of edge-on LSBGs. It is worth noting that the surface brightness and axis ratio of some detected samples slightly exceed the predefined range for edge-on LSBGs, which represents a precision limitation of the model. Nonetheless, our model significantly enhances the level of automation in the recognition of edge-on LSBGs, reducing the need for manual intervention.

\subsection{The Catalog of the Candidate LSBGs}

To ensure the production of a more dependable catalog, we implemented selection criteria based on axis ratio and central surface brightness ($expAB\_g$ $<$ 0.3, $\mu_{0,B}$ $>$ 22 $mag \ arcsec^{-2}$). In addition to this, we removed three cosmic-ray contaminations through visual inspection, resulting in a final count of 40,759 edge-on LSBG candidates.  Among them, 40,558 candidate LSBGs correspond to SDSS galaxies, while 201 candidates are newly detected sources not present in the SDSS Galaxy view. A portion of the catalog is shown in Table \ref{tbl2} and the full version is available online at \url{https://github.com/worldoutside/Edge-on_LSBG}.

\begin{table}[htbp]
 \centering
    \caption{Catalog of Candidate LSBGs Detected in SDSS DR16, Sorted by Model-predicted Confidence}
\label{tbl2} 
\setlength{\tabcolsep}{1mm}{
\begin{tabular}{ccccc}
\hline
\hline
R.A.      & Decl.      & Model-predicted & Anomaly & Flag$^{a}$ \\
(J2000)   & (J2000)    & Confidence      & Score   &      \\ \hline

147.237788 & 33.043279 & 0.8819 & 0.1853 & 0 \\
351.174469 & -2.791721 & 0.8775 & 0.1289 & 0 \\
19.768418 & -0.138547 &	0.8717 & 0.3517  & 0 \\
210.675864 & 9.164149 &	0.8687 & 0.4142 & 0\\
332.464216 & 7.429644 &	0.8680 & 0.4868 & 0 \\
202.637660 & -1.659615 & 0.8639 & 0.2724 & 0 \\
190.066722 & -7.682345 & 0.8622 & 0.2797 &	0 \\
204.009414 & 8.186349 & 0.8603 & 0.3557 & 0 \\ 
211.839267 & 55.491370 & 0.8592 & 0.1170 & 0 \\ 
228.918833 & 8.217672 &	0.8583 & 0.2326	& 0 \\

233.022783 & 41.024649 & 0.8468 & 0.1069 & 1 \\
326.776630  &  24.334284  & 0.8442 & 0.2621 & 1 \\
215.428509 & 33.417431 & 0.8186 & 0.0863 & 1 \\
253.316873 & 47.226675 & 0.8135 &0.1242 & 1 \\
114.417771 & 22.441127 & 0.8128 & 0.0978 & 1 \\
2.578499 & 29.137902 & 0.8122 & 0.0783 & 1 \\
319.653219 & 23.389592 & 0.8086 & 0.2853 & 1 \\
217.701806 & 53.344508 & 0.8074 & 0.0963 & 1 \\ 
228.508681 & 38.256345 & 0.803 & 0.0596 & 1 \\
2.662196 & -19.773956 & 0.800 & 0.0601 & 1 \\

\hline

\end{tabular}}

\begin{threeparttable}
\begin{tablenotes}
    \footnotesize
    \item\textbf{Notes.}
    A copy of the catalog is also available at \url{https://github.com/worldoutside/Edge-on_LSBG} (the full catalog includes 40,759 candidate LSBGs).
    
\item[a] Flag is used to denote the source of R.A. and Decl. When Flag=0, the first two columns correspond to the SDSS R.A. and SDSS Decl. When Flag=1, the first two columns represent the model-predicted R.A. and Decl.
\end{tablenotes}
\end{threeparttable}
\end{table}

\section{Conclusion} \label{sec:conclusion}

In this paper, we present an edge-on LSBG catalog identified from SDSS DR16 field images using deep-learning methods. With a sample of 281 edge-on HI-rich LSBGs from \citet{he2020sample}, a deep-learning object detection model was built using the YOLOv5 algorithm, achieving a recall of 94.64\% and a purity of 95.38\% for the test set. We then applied the model to search for edge-on LSBGs from 938,046 composite images in SDSS DR16, and 52,293 candidate LSBGs were identified. Subsequently, the candidate LSBGs were purified by the Deep-SVDD anomaly detection model. We showed the properties of the sample of candidate LSBGs, including the $B$-band central surface brightness, axis ratio, and color-magnitude relation. The properties of the detected samples are in good agreement with those of the training sample.
Finally, we provided a sample that includes 40,759 candidate edge-on LSBGs, which is a wide-area sample for future studies investigating the properties of edge-on LSBGs within the realm of galaxy research.

This study utilizes deep-learning methods for the automatic detection of edge-on LSBGs, leading to a significant enhancement in the automation of LSBG detection while reducing the need for manual inspection. This approach remains effective in identifying sources that are challenging to extract parameters from using traditional methods, including relatively dim and irregular galaxies. Importantly, the established detection model operates independently of photometric parameters, enabling the identification of sources that might be missed by photometric methods. This technology holds promise for the development of intelligent image analysis tools to support future large-scale sky surveys.

\section{Acknowledgments}
This study was supported by Shandong Province Natural Science Foundation grant No. ZR2022MA089, the National Natural Science Foundation of China (NSFC) grant Nos. U1931209, and 11803016, and the Chinese Space Station Telescope project. D.W. is supported by the National Natural Science Foundation of China (NSFC) grant Nos. U1931109 and 11733006 and the Youth Innovation Promotion Association, Chinese Academy of Sciences (CAS), No. 2020057. Y.B. is supported by the National Natural Science Foundation of China under grant No. 11873037 and partially supported by the Young Scholars Program of Shandong University, Weihai (2016WHWLJH09), and the science research grants from the China Manned Space Project with Nos. CMS-CSST-2021-B05 and CMS-CSST-2021-A08.
 
We thank the SDSS team for the released SDSS images and parameter catalog. Funding for the SDSS has been provided by the Alfred P. Sloan Foundation, the Participating Institutions, the National Science Foundation, the US Department of Energy, NASA, the Japanese Monbukagakusho, the Max Planck Society, and the Higher Education Funding Council for England. The SDSS Web Site is \url{http://www.sdss.org}.

\software{astropy \citep{robitaille2013astropy, price2018astropy},  
          matplotlib \citep{hunter2007matplotlib},
          numpy \citep{van2011numpy},
          pandas \citep{mckinney2010Data},
          python3 \citep{van2009python},
          pytorch \citep{paszke2019pytorch}
          }

\bibliography{references-quoted}{}

\begin{thebibliography}{}
\expandafter\ifx\csname natexlab\endcsname\relax\def\natexlab#1{#1}\fi
\providecommand{\url}[1]{\href{#1}{#1}}
\providecommand{\dodoi}[1]{doi:~\href{http://doi.org/#1}{\nolinkurl{#1}}}
\providecommand{\doeprint}[1]{\href{http://ascl.net/#1}{\nolinkurl{http://ascl.net/#1}}}
\providecommand{\doarXiv}[1]{\href{https://arxiv.org/abs/#1}{\nolinkurl{https://arxiv.org/abs/#1}}}

\bibitem[{Abazajian {et~al.}(2009)Abazajian, Adelman-McCarthy, Ag{\"u}eros,
  Allam, Prieto, An, Anderson, Anderson, Annis, Bahcall,
  {et~al.}}]{abazajian2009seventh}
Abazajian, K.~N., Adelman-McCarthy, J.~K., Ag{\"u}eros, M.~A., {et~al.} 2009,
  The Astrophysical Journal Supplement Series, 182, 543,
  \dodoi{10.1088/0067-0049/182/2/543}

\bibitem[{Adami {et~al.}(2006)Adami, Scheidegger, Ulmer, Durret, Mazure, West,
  Conselice, Gregg, Kasun, Pell{\'o}, {et~al.}}]{adami2006deep}
Adami, C., Scheidegger, R., Ulmer, M., {et~al.} 2006, Astronomy \&
  Astrophysics, 459, 679, \dodoi{10.1051/0004-6361:20053758}

\bibitem[{Adams {et~al.}(2011)Adams, Uson, Hill, \& MacQueen}]{adams2011new}
Adams, J.~J., Uson, J.~M., Hill, G.~J., \& MacQueen, P.~J. 2011, The
  Astrophysical Journal, 728, 107, \dodoi{10.1088/0004-637X/728/2/107}

\bibitem[{Ahumada {et~al.}(2020)Ahumada, Prieto, Almeida, Anders, Anderson,
  Andrews, Anguiano, Arcodia, Armengaud, Aubert, {et~al.}}]{ahumada202016th}
Ahumada, R., Prieto, C.~A., Almeida, A., {et~al.} 2020, The Astrophysical
  Journal Supplement Series, 249, 3, \dodoi{10.3847/1538-4365/ab929e}

\bibitem[{Bergvall {et~al.}(2010)Bergvall, Zackrisson, \&
  Caldwell}]{bergvall2010red}
Bergvall, N., Zackrisson, E., \& Caldwell, B. 2010, Monthly Notices of the
  Royal Astronomical Society, 405, 2697,
  \dodoi{10.1111/j.1365-2966.2010.16650.x}

\bibitem[{Bernardi {et~al.}(2010)Bernardi, Shankar, Hyde, Mei, Marulli, \&
  Sheth}]{bernardi2010galaxy}
Bernardi, M., Shankar, F., Hyde, J., {et~al.} 2010, Monthly Notices of the
  Royal Astronomical Society, 404, 2087,
  \dodoi{10.1111/j.1365-2966.2010.16425.x}

\bibitem[{Bizyaev \& Kajsin(2004)}]{bizyaev2004stellar}
Bizyaev, D., \& Kajsin, S. 2004, The Astrophysical Journal, 613, 886,
  \dodoi{10.1086/423229}

\bibitem[{Bizyaev {et~al.}(2014)Bizyaev, Kautsch, Mosenkov, Reshetnikov,
  Sotnikova, Yablokova, \& Hillyer}]{bizyaev2014catalog}
Bizyaev, D., Kautsch, S., Mosenkov, A., {et~al.} 2014, The Astrophysical
  Journal, 787, 24, \dodoi{10.1088/0004-637X/787/1/24}

\bibitem[{Bizyaev {et~al.}(2017)Bizyaev, Kautsch, Sotnikova, Reshetnikov, \&
  Mosenkov}]{bizyaev2017very}
Bizyaev, D., Kautsch, S., Sotnikova, N.~Y., Reshetnikov, V.~P., \& Mosenkov,
  A.~V. 2017, Monthly Notices of the Royal Astronomical Society, 465, 3784,
  \dodoi{10.1093/mnras/stw2972}

\bibitem[{Bochkovskiy {et~al.}(2020)Bochkovskiy, Wang, \&
  Liao}]{bochkovskiy2020yolov4}
Bochkovskiy, A., Wang, C.-Y., \& Liao, H.-Y.~M. 2020, arXiv preprint
  arXiv:2004.10934, \dodoi{10.48550/arXiv.2004.10934}

\bibitem[{Breiman(2001)}]{breiman2001random}
Breiman, L. 2001, Machine learning, 45, 5, \dodoi{10.1023/A:1010933404324}

\bibitem[{Caldwell \& Bergvall(2006)}]{2006Edge}
Caldwell, B., \& Bergvall, N. 2006, Proceedings of the International
  Astronomical Union, 2, 82, \dodoi{10.1017/S1743921306005199}

\bibitem[{Ceccarelli {et~al.}(2012)Ceccarelli, Herrera-Camus, Lambas, Galaz, \&
  Padilla}]{ceccarelli2012low}
Ceccarelli, L., Herrera-Camus, R., Lambas, D., Galaz, G., \& Padilla, N. 2012,
  Monthly Notices of the Royal Astronomical Society: Letters, 426, L6,
  \dodoi{10.1111/j.1745-3933.2012.01311.x}

\bibitem[{Chandola {et~al.}(2009)Chandola, Banerjee, \&
  Kumar}]{chandola2009anomaly}
Chandola, V., Banerjee, A., \& Kumar, V. 2009, ACM computing surveys (CSUR),
  41, 1, \dodoi{10.1145/1541880.1541882}

\bibitem[{De~Grijs(1998)}]{de1998global}
De~Grijs, R. 1998, Monthly Notices of the Royal Astronomical Society, 299, 595,
  \dodoi{10.1046/j.1365-8711.1998.01896.x}

\bibitem[{D{\'\i}az-Garc{\'\i}a {et~al.}(2022)D{\'\i}az-Garc{\'\i}a,
  Comer{\'o}n, Courteau, Watkins, Knapen, \& Rom{\'a}n}]{diaz2022linking}
D{\'\i}az-Garc{\'\i}a, S., Comer{\'o}n, S., Courteau, S., {et~al.} 2022,
  Astronomy \& Astrophysics, 667, A109, \dodoi{10.1051/0004-6361/202142447}

\bibitem[{Du {et~al.}(2015)Du, Wu, Lam, Zhu, Lei, \& Zhou}]{du2015low}
Du, W., Wu, H., Lam, M.~I., {et~al.} 2015, The Astronomical Journal, 149, 199,
  \dodoi{10.1088/0004-6256/149/6/199}

\bibitem[{Du {et~al.}(2017)Du, Wu, Zhu, Zheng, \& Filippenko}]{du2017long}
Du, W., Wu, H., Zhu, Y., Zheng, W., \& Filippenko, A.~V. 2017, The
  Astrophysical Journal, 837, 152, \dodoi{10.3847/1538-4357/aa6194}

\bibitem[{DuToit {et~al.}(2012)DuToit, Steyn, \& Stumpf}]{dutoit2012graphical}
DuToit, S.~H., Steyn, A. G.~W., \& Stumpf, R.~H. 2012, Graphical exploratory
  data analysis (Springer Science \& Business Media)

\bibitem[{Gerritsen \& de~Blok(1999)}]{gerritsen1999star}
Gerritsen, J.~P., \& de~Blok, W. 1999, Astronomy and Astrophysics, 342, 655,
  \dodoi{10.48550/arXiv.astro-ph/9810096}

\bibitem[{Giovanelli(2007)}]{giovanelli2007alfalfa}
Giovanelli, R. 2007, Nuovo Cimento B Serie, 122, 1097,
  \dodoi{10.1393/ncb/i2008-10442-9}

\bibitem[{Greco {et~al.}(2018)Greco, Greene, Strauss, Macarthur, Flowers,
  Goulding, Huang, Kim, Komiyama, Leauthaud, Leisman, Lupton, Sif{\'{o}}n, \&
  Wang}]{Greco2018}
Greco, J.~P., Greene, J.~E., Strauss, M.~A., {et~al.} 2018, The Astrophysical
  Journal, 857, 104, \dodoi{10.3847/1538-4357/aab842}

\bibitem[{He {et~al.}(2020)He, Wu, Du, Liu, Lei, Zhao, \& Zhang}]{he2020sample}
He, M., Wu, H., Du, W., {et~al.} 2020, The Astrophysical Journal Supplement
  Series, 248, 33, \dodoi{10.3847/1538-4365/ab8ead}

\bibitem[{He {et~al.}(2019)He, Wu, Du, Wicker, Zhao, Lei, \& Liu}]{he2019edge}
---. 2019, The Astrophysical Journal, 880, 30, \dodoi{10.3847/1538-4357/ab2710}

\bibitem[{Hunter(2007)}]{hunter2007matplotlib}
Hunter, J.~D. 2007, Computing in science \& engineering, 9, 90,
  \dodoi{10.1109/MCSE.2007.55}

\bibitem[{Impey \& Bothun(1997)}]{impey1997low}
Impey, C., \& Bothun, G. 1997, Annual Review of Astronomy and Astrophysics, 35,
  267, \dodoi{10.1146/annurev.astro.35.1.267}

\bibitem[{Impey {et~al.}(2001)Impey, Burkholder, \& Sprayberry}]{impey2001high}
Impey, C., Burkholder, V., \& Sprayberry, D. 2001, The Astronomical Journal,
  122, 2341, \dodoi{10.1086/323537}

\bibitem[{Karachentsev {et~al.}(1999)Karachentsev, Karachentseva, Kudrya,
  Sharina, \& Parnovskij}]{karachentsev1999revised}
Karachentsev, I., Karachentseva, V., Kudrya, Y.~N., Sharina, M., \& Parnovskij,
  S. 1999, Bulletin of the Special Astrophysics Observatory, 47, 5,
  \dodoi{10.48550/arXiv.astro-ph/0305566}

\bibitem[{Kingma \& Ba(2014)}]{kingma2014adam}
Kingma, D.~P., \& Ba, J. 2014, arXiv preprint arXiv:1412.6980,
  \dodoi{10.48550/arXiv.1412.6980}

\bibitem[{LeCun {et~al.}(1998)LeCun, Bottou, Bengio, \&
  Haffner}]{lecun1998gradient}
LeCun, Y., Bottou, L., Bengio, Y., \& Haffner, P. 1998, Proceedings of the
  IEEE, 86, 2278, \dodoi{10.1109/5.726791}

\bibitem[{Liang {et~al.}(2007)Liang, Hu, Liu, \& Liu}]{liang2007sdss}
Liang, Y., Hu, J., Liu, F., \& Liu, Z. 2007, The Astronomical Journal, 134,
  759, \dodoi{10.1086/519957}

\bibitem[{MacLachlan {et~al.}(2011)MacLachlan, Matthews, Wood, \&
  Gallagher}]{maclachlan2011stability}
MacLachlan, J.~M., Matthews, L.~D., Wood, K., \& Gallagher, J. 2011, The
  Astrophysical Journal, 741, 6, \dodoi{10.1088/0004-637X/741/1/6}

\bibitem[{Masci {et~al.}(2011)Masci, Meier, Cire{\c{s}}an, \&
  Schmidhuber}]{masci2011stacked}
Masci, J., Meier, U., Cire{\c{s}}an, D., \& Schmidhuber, J. 2011, in
  International conference on artificial neural networks, Springer, 52--59

\bibitem[{Matthews \& De~Grijs(2004)}]{matthews2004optical}
Matthews, L., \& De~Grijs, R. 2004, The Astronomical Journal, 128, 137,
  \dodoi{10.1086/421363}

\bibitem[{Matthews {et~al.}(1999)Matthews, Gallagher~III, \&
  Van~Driel}]{matthews1999extraordinary}
Matthews, L., Gallagher~III, J., \& Van~Driel, W. 1999, The Astronomical
  Journal, 118, 2751, \dodoi{10.1086/301128}

\bibitem[{Matthews \& Gao(2001)}]{matthews2001co}
Matthews, L., \& Gao, Y. 2001, The Astrophysical Journal, 549, L191,
  \dodoi{10.1086/319175}

\bibitem[{Matthews {et~al.}(2005)Matthews, Gao, Uson, \&
  Combes}]{matthews2005detections}
Matthews, L.~D., Gao, Y., Uson, J.~M., \& Combes, F. 2005, The Astronomical
  Journal, 129, 1849, \dodoi{10.1086/428857}

\bibitem[{Matthews \& Uson(2008)}]{matthews2008hi}
Matthews, L.~D., \& Uson, J.~M. 2008, The Astronomical Journal, 135, 291,
  \dodoi{10.1088/0004-6256/135/1/291}

\bibitem[{Matthews \& Wood(2003)}]{matthews2003high}
Matthews, L.~D., \& Wood, K. 2003, The Astrophysical Journal, 593, 721,
  \dodoi{10.1086/376602}

\bibitem[{Mcgaugh {et~al.}(1995)Mcgaugh, Schombert, \&
  Bothun}]{mcgaugh1995morphology}
Mcgaugh, S., Schombert, J., \& Bothun, G. 1995, Astronomical Journal, 109,
  2019, \dodoi{10.1086/117427}

\bibitem[{Mckinney(2010)}]{mckinney2010Data}
Mckinney, W. 2010, proc.python sci.conf, \dodoi{{10.25080/MAJORA-92BF1922-00A}}

\bibitem[{Narayanan \& Banerjee(2022)}]{narayanan2022superthin}
Narayanan, G., \& Banerjee, A. 2022, Monthly Notices of the Royal Astronomical
  Society, 514, 5126, \dodoi{10.1093/mnras/stac1662}

\bibitem[{Neeser {et~al.}(2002)Neeser, Sackett, De~Marchi, \&
  Paresce}]{neeser2002detection}
Neeser, M.~J., Sackett, P.~D., De~Marchi, G., \& Paresce, F. 2002, Astronomy \&
  Astrophysics, 383, 472, \dodoi{10.1051/0004-6361:20011757}

\bibitem[{Paszke {et~al.}(2019)Paszke, Gross, Massa, Lerer, Bradbury, Chanan,
  Killeen, Lin, Gimelshein, Antiga, {et~al.}}]{paszke2019pytorch}
Paszke, A., Gross, S., Massa, F., {et~al.} 2019, Advances in neural information
  processing systems, 32, \dodoi{10.48550/arXiv.1912.01703}

\bibitem[{Platt(1998)}]{platt1998sequential}
Platt, J.~C. 1998, Microsoft Research, \dodoi{US4558132 A}

\bibitem[{Pohlen {et~al.}(2003)Pohlen, Balcells, L{\"u}tticke, \&
  Dettmar}]{pohlen2003evidence}
Pohlen, M., Balcells, M., L{\"u}tticke, R., \& Dettmar, R.-J. 2003, Astronomy
  \& Astrophysics, 409, 485, \dodoi{10.1051/0004-6361:20031091}

\bibitem[{Price-Whelan {et~al.}(2018)Price-Whelan, Sip{\H{o}}cz, G{\"u}nther,
  Lim, Crawford, Conseil, Shupe, Craig, Dencheva, Ginsburg,
  {et~al.}}]{price2018astropy}
Price-Whelan, A.~M., Sip{\H{o}}cz, B., G{\"u}nther, H., {et~al.} 2018, The
  Astronomical Journal, 156, 123, \dodoi{10.3847/1538-3881/aabc4f}

\bibitem[{Pustilnik {et~al.}(2010)Pustilnik, Tepliakova, Kniazev, Martin, \&
  Burenkov}]{pustilnik2010sdss}
Pustilnik, S., Tepliakova, A., Kniazev, A., Martin, J.-M., \& Burenkov, A.
  2010, Monthly Notices of the Royal Astronomical Society, 401, 333,
  \dodoi{10.1111/j.1365-2966.2009.15637.x}

\bibitem[{Redmon {et~al.}(2016)Redmon, Divvala, Girshick, \&
  Farhadi}]{redmon2016you}
Redmon, J., Divvala, S., Girshick, R., \& Farhadi, A. 2016, in Proceedings of
  the IEEE conference on computer vision and pattern recognition, 779--788,
  \dodoi{10.1109/CVPR.2016.91}

\bibitem[{Redmon \& Farhadi(2017)}]{redmon2017yolo9000}
Redmon, J., \& Farhadi, A. 2017, in Proceedings of the IEEE conference on
  computer vision and pattern recognition, 7263--7271,
  \dodoi{10.1109/CVPR.2017.690}

\bibitem[{Redmon \& Farhadi(2018)}]{redmon2018yolov3}
Redmon, J., \& Farhadi, A. 2018, arXiv preprint arXiv:1804.02767,
  \dodoi{10.48550/arXiv.1804.02767}

\bibitem[{Robitaille {et~al.}(2013)Robitaille, Tollerud, Greenfield,
  Droettboom, Bray, Aldcroft, Davis, Ginsburg, Price-Whelan, Kerzendorf,
  {et~al.}}]{robitaille2013astropy}
Robitaille, T.~P., Tollerud, E.~J., Greenfield, P., {et~al.} 2013, Astronomy \&
  Astrophysics, 558, A33, \dodoi{10.1051/0004-6361/201322068}

\bibitem[{Ruff {et~al.}(2018)Ruff, Vandermeulen, Goernitz, Deecke, Siddiqui,
  Binder, M{\"u}ller, \& Kloft}]{ruff2018deep}
Ruff, L., Vandermeulen, R., Goernitz, N., {et~al.} 2018, in International
  conference on machine learning, PMLR, 4393--4402

\bibitem[{Sarkar \& Jog(2019)}]{sarkar2019flaring}
Sarkar, S., \& Jog, C.~J. 2019, Astronomy \& Astrophysics, 628, A58,
  \dodoi{10.1051/0004-6361/201935430}

\bibitem[{Tanoglidis {et~al.}(2021{\natexlab{a}})Tanoglidis,
  {\'C}iprijanovi{\'c}, \& Drlica-Wagner}]{tanoglidis2021deepshadows}
Tanoglidis, D., {\'C}iprijanovi{\'c}, A., \& Drlica-Wagner, A.
  2021{\natexlab{a}}, Astronomy and Computing, 35, 100469,
  \dodoi{10.1016/j.ascom.2021.100469}

\bibitem[{Tanoglidis {et~al.}(2021{\natexlab{b}})Tanoglidis, Drlica-Wagner,
  Wei, Li, S{\'a}nchez, Zhang, Peter, Feldmeier-Krause, Prat, Casey,
  {et~al.}}]{tanoglidis2021shadows}
Tanoglidis, D., Drlica-Wagner, A., Wei, K., {et~al.} 2021{\natexlab{b}}, The
  Astrophysical Journal Supplement Series, 252, 18,
  \dodoi{10.3847/1538-4365/abca89}

\bibitem[{Tukey {et~al.}(1977)}]{tukey1977exploratory}
Tukey, J.~W., {et~al.} 1977, Exploratory data analysis, Vol.~2 (Reading, MA),
  \dodoi{10.2307/2286300}

\bibitem[{Van Der~Kruit {et~al.}(2001)Van Der~Kruit, Jim{\'e}nez-Vicente,
  Kregel, \& Freeman}]{van2001kinematics}
Van Der~Kruit, P., Jim{\'e}nez-Vicente, J., Kregel, M., \& Freeman, K. 2001,
  Astronomy \& Astrophysics, 379, 374, \dodoi{10.1051/0004-6361:20011311}

\bibitem[{Van Der~Walt {et~al.}(2011)Van Der~Walt, Colbert, \&
  Varoquaux}]{van2011numpy}
Van Der~Walt, S., Colbert, S.~C., \& Varoquaux, G. 2011, Computing in science
  \& engineering, 13, 22, \dodoi{10.1109/MCSE.2011.37}

\bibitem[{Van~Rossum \& Drake(2009)}]{van2009python}
Van~Rossum, G., \& Drake, F.~L. 2009, Python/C Api Manual-Python 3
  (CreateSpace)

\bibitem[{Yi {et~al.}(2022)Yi, Li, Du, Liu, Liang, Xing, Pan, Bu, Kong, \&
  Wu}]{yi2022automatic}
Yi, Z., Li, J., Du, W., {et~al.} 2022, Monthly Notices of the Royal
  Astronomical Society, 513, 3972, \dodoi{10.1093/mnras/stac775}

\end{thebibliography}
\bibliographystyle{aasjournal}
\end{document}